\documentclass[aps,prb,twocolumn,superscriptaddress,longbibliography]{revtex4-2}
\usepackage{graphicx}
\usepackage{amssymb,amsmath}
\usepackage{bm}
\usepackage{dcolumn,dsfont}
\usepackage{subfigure}
\usepackage{float}
\usepackage[OT1]{fontenc} 
\usepackage{url}
\usepackage{mathrsfs}
\usepackage{slashed}
\usepackage{color}
\usepackage{verbatim}
\usepackage{enumitem}
\usepackage{txfonts}
\usepackage{soul,physics}
\usepackage[driverfallback=dvipdfm]{hyperref}
\hypersetup{pdfpagemode=FullScreen,colorlinks=true,breaklinks,urlcolor=blue,linkcolor=blue,citecolor=blue}
\usepackage{natbib}
\usepackage{environ}

\usepackage{subfigure}
\usepackage{comment}
\usepackage{hyperref}
\usepackage{amssymb}
\usepackage{bm}
\usepackage{graphicx}
\usepackage{amsmath,amssymb}
\usepackage{tikz,fp}
\usepackage{tikz-cd}

\pgfdeclarepatternformonly{south west lines}{\pgfqpoint{-0pt}{-0pt}}{\pgfqpoint{3pt}{3pt}}{\pgfqpoint{3pt}{3pt}}{
	\pgfsetlinewidth{0.4pt}
	\pgfpathmoveto{\pgfqpoint{0pt}{0pt}}
	\pgfpathlineto{\pgfqpoint{3pt}{3pt}}
	\pgfpathmoveto{\pgfqpoint{2.8pt}{-.2pt}}
	\pgfpathlineto{\pgfqpoint{3.2pt}{.2pt}}
	\pgfpathmoveto{\pgfqpoint{-.2pt}{2.8pt}}
	\pgfpathlineto{\pgfqpoint{.2pt}{3.2pt}}
	\pgfusepath{stroke}}

\tikzset{
	mid arrow/.style={postaction={decorate,decoration={
				markings,
				mark=at position .575 with {\arrow{stealth}}
	}}},
	near arrow/.style={postaction={decorate,decoration={
				markings,
				mark=at position .275 with {\arrow{stealth}}
	}}},
	far arrow/.style={postaction={decorate,decoration={
				markings,
				mark=at position .800 with {\arrow{stealth}}
	}}},
	snake arrow/.style={fixed point arithmetic, decorate, decoration={snake,amplitude=2pt, segment length=11pt},postaction={decoration={markings,mark=at position 0.625 with {\arrow{stealth}}},decorate}},
}
\tikzset{
  baseline = -0.5ex,
  wavy/.style = {
    thick,
    decorate,
    decoration={snake,amplitude=2pt,segment length=5pt}},
  sdot/.style = {
    circle,
    draw=none,
    fill=black,
    minimum size=2.5pt,
    inner sep=0pt},
  bdot/.style = {
    circle,
    draw=none,
    fill=black,
    minimum size=4pt,
    inner sep=0pt},
  svertex/.style = {
    circle,
    draw=black,
    thick,
    fill=lightgray,
    minimum size=14pt,
    inner sep=1pt},
  bvertex/.style = {
    circle,
    draw=black,
    thick,
    fill=lightgray,
    minimum size=24pt},
  bvertexsmall/.style = {
    circle,
    draw=black,
    thick,
    fill=lightgray,
    minimum size=7pt},
  bvertexnormal/.style = {
    circle,
    draw=black,
    thick,
    fill=lightgray,
    minimum size=16pt},
    bvertexnormal2/.style = {
    circle,
    draw=black,
    thick,
    fill=lightgray,
    minimum size=24pt},
  dvertex/.style = {
    circle,
    draw=black,
    thick,
    fill=gray,
    minimum size=25pt}}

\NewEnviron{myequation}{%
    \begin{equation}
    \scalebox{0.91}{$\displaystyle{\BODY}$}
    \end{equation}
    }

\begin{document}

\preprint{APS/123-QED}

\title{Validity of the Lieb–Schultz–Mattis Theorem in Long-Range Interacting Systems}% Force line breaks with \\

\author{Yi-Neng Zhou}
\email{zhouyn.physics@gmail.com}
\affiliation{Institute for Advanced Study, Tsinghua University, Beijing 100084, China}
\author{Xingyu Li}
\email{xyli22@mails.tsinghua.edu.cn}
\affiliation{Institute for Advanced Study, Tsinghua University, Beijing 100084, China}

% \collaboration{MUSO Collaboration}%\noaffiliation

\date{\today}% It is always \today, today,

\begin{abstract}
The Lieb-Schultz-Mattis (LSM) theorem asserts that microscopic details of the system can impose non-trivial constraints on the system's low-energy properties.  While traditionally applied to short-range interaction systems, where locality ensures a vanishing spectral gap in large system size limit, the impact of long-range interactions on the LSM theorem remains an open question. Long-range interactions are prevalent in experimental platforms such as Rydberg atoms, dipolar quantum gases, polar molecules, optical cavities, and trapped ions, where the interaction decay exponent can be experimentally tuned. We extend the LSM theorem in one dimension to long-range interacting systems and find that the LSM theorem holds for exponentially or power-law two-body interactions with a decay exponent $\alpha > 2$. However, for power-law interactions with $\alpha < 2$, the constraints of the LSM theorem on the ground state do not apply. Numerical simulations of long-range versions of the Heisenberg and Majumdar–Ghosh models, both satisfying the LSM symmetry requirements, are also provided. Our results suggest promising directions for experimental validation of the LSM theorem in systems with tunable long-range interactions.
\end{abstract}

\maketitle

%\tableofcontents

\section{Introduction}
\label{sec:intro}
% TODO: write your article here.
The celebrated Lieb--Schultz--Mattis (LSM) theorem was first proved by Lieb, Schultz, and Mattis in 1961~\cite{Lieb1961}. They demonstrated that a one-dimensional quantum spin system with translational symmetry, SO(3) rotational symmetry, and half-odd integer spin per unit cell cannot have a unique, gapped ground state\cite{Lieb1961,Affleck1986,Oshikawa2000, Hastings2004}. This theorem establishes an ultraviolet-infrared (UV-IR) correspondence in the quantum system. It indicates that microscopic details of the system (such as whether the spin of each site is a half-odd integer) can impose non-trivial constraints on the system's low-energy properties (such as the ground state). The most critical step in proving the LSM theorem is in showing that, in a system with translational and rotational symmetry and half-odd-integer spin per unit cell, the translation operator $\hat{T}$ anti-commutes with the spin twist operator $\hat{U}^{\text{twist}}$. This anti-commutation relation depends only on the properties of the Hilbert space and not on the specific form of the Hamiltonian. 
The original LSM theorem has various generalizations, including extensions to higher dimensions \cite{Oshikawa2000, Hastings2004}, more sophisticated on-site and space groups \cite{chen2011classification, Parameswaran13, Watanabe_2015, Po_2017}, symmetry-protected and enriched topological systems \cite{Zaletel_2015, Cheng2016, Cho2017, huang2017buildinga, lu2017liebschultzmattis, Yang_2018, jian2018liebschultzmattis, Metlitski2018, Ogata2019, Else2020, Lu_2020, jiang2021generalizeda, ye2022topological}, fermionic systems\cite{fidkowski2018surface,Cheng_2019,zheng2022unconventional,Aksoy_2021} and open systems\cite{zhouRevivingLiebSchultzMattisTheorem2023a, Kawabata2024}. 

In the original LSM theorem, short-range interaction systems are typically considered, where the good locality of the Hamiltonian makes it straightforward to prove that its spectral gap vanishes in the large system size limit. 
In real experiments, the presence of long-range interactions in quantum systems induces several
peculiar features in their equilibrium and out-of-equilibrium behavior~\cite{defenuLongrangeInteractingQuantum2023}, which makes it natural to wonder the validity of the LSM theorem in systems with long-range interactions. Meanwhile, the long-range quantum systems are currently being realized in several experimental platforms such as Rydberg atoms~\cite{isenhowerDemonstrationNeutralAtom2010}, dipolar quantum gases~\cite{Lahaye2009}, polar molecules~\cite{Carr_2009}, quantum gases coupled to optical cavities~\cite{Mivehvar_2021,ritschColdAtomsCavitygenerated2013}, and trapped ions~\cite{blattQuantumSimulationsTrapped2012,Schneider2012ExperimentalQS,Monroe2021}, where the decaying exponent of the long-range interactions is a tunable experimental parameter in some of the experimental platforms. Therefore, it will be promising to test the validity of the LSM theorem in real experiments in the future.

In this work, we provide a detailed derivation of the LSM theorem for long-range systems with two-body interactions. We find that for exponentially decaying interactions or power-law interactions with a decay exponent $\alpha > 2$, the LSM theorem still holds. However, for power-law interactions with a decay exponent $\alpha < 2$, the LSM theorem's constraints on the ground state may fail. We also numerically simulate two long-range generalizations of familiar models, the Heisenberg model and the Majumdar–Ghosh model, which satisfy the symmetry requirements of the LSM theorem. Our simulations demonstrate a possible transition from a gapless (degenerate) phase to a uniquely gapped phase when decay exponent $\alpha$ crosses 2 with open boundary conditions.

\section{Review of the proof of the original LSM theorem}

Let us first review the proof of the original LSM theorem. We will then see how to modify this derivation to fit into the long-range interaction system in the next section. This straightforward proof originates from the original paper on the LSM theorem~\cite{Lieb1961}. In 1961, Lieb, Schultz, and Mattis proved that a one-dimensional periodic chain of length $L$, with translational symmetry and SO(3) spin rotation symmetry, with half-integer spin per unit cell, has an excitation gap bounded by $\text{const}/L$.

The simple proof is as follows\cite{Affleck1988}:
\begin{itemize}
	\item{1.} We assume the system has a unique ground state $|\psi_0\rangle$, and then construct a low-energy excited state $|\psi_1\rangle$ by twisting the boundary condition of the ground state.
	\item{2.} We prove that this excited state $|\psi_1\rangle$ is orthogonal to $|\psi_0\rangle$.
	\item{3.} 
	We prove the excitation energy of excited state $|\psi_1\rangle$ is proportional to $\text{const}/L$ which vanishes in the $L \to \infty$ limit. This shows that the system cannot have a unique gapped ground state.
\end{itemize}

We take the one-dimensional spin-$\frac{1}{2}$ antiferromagnetic Heisenberg model as an example. Its Hamiltonian is 
\begin{equation}
	\hat{H}=J\sum_{j=1}^L \hat{\bf S}_j\cdot \hat{\bf S}_{j+1}.
\end{equation}
Here $j$ is the site index, and we set the total length of the system to $L$. Because LSM requires the system to be translational invariant, we use the periodic boundary condition: $\hat{\bf S}_{L+1}=\hat{\bf S}_1$. We can easily see that this model satisfies the translation invariance, SO(3) rotation invariance required by the LSM theorem, and the condition that the spin on each site is a half-odd integer.

We first define the twist operator required in the first step of the proof:
\begin{equation}
	\hat{U}^{\text{twist}}=\exp[i\frac{2\pi }{L}\sum_{j-1}^L j\hat{\bf S}_j^z].
	\label{twist}
\end{equation}
This twist operator is as a rotation operation around the z-axis for each site of the system, where the rotation angle is proportional to the site index. For the spin on each site of this one-dimensional chain, the rotation angle gradually increases from $0$ to $2\pi$. This twist operator can also be interpreted as threading a unit flux (of the $U(1)$ rotation symmetry around the z-axis) through the system, causing the wave function of this one-dimensional chain to acquire an additional position-dependent phase factor. We construct an excited state $|\psi_1\rangle$ by applying the twist operator to the ground state $|\psi_0\rangle$:
\begin{equation}
	|\psi_1\rangle=\hat{U}^{\text{twist}}|\psi_0\rangle.
\end{equation}

Then we prove  $|\psi_1\rangle$ has momentum $\pi$ relative to the ground state $|\psi_0\rangle$. Since $|\psi_0\rangle$ is assumed to be the unique ground state of $\hat{H}$ which has translational symmetry, $|\psi_1\rangle$ has to be orthogonal to $|\psi_0\rangle$. To do this, we compute how $\hat{U}^{\mathrm{twist}}$ transforms under the translation operation
\begin{equation} 
    \begin{split}
		\hat{T}\hat{U}^{\mathrm{twist}}\hat{T}^{-1}=&\exp\lbrace
  i\frac{2\pi}{L}[\sum_{n=2}^L (n-1)\hat{\bf S}_{n+1}^z+L\hat{\bf S}_1^z]\rbrace \\
  =&\hat{U}^{\mathrm{twist}}\exp[
  -i\frac{2\pi}{L}\sum_{n=1}^L \hat{\bf S}_{n+1}^z] \exp[i2\pi \hat{\bf S}_1^z].
	\end{split}
 \label{eq:twist}
\end{equation}
Here $\hat{T}$ is the translation operator that moves the state by one site. The last factor $\exp[i2\pi \hat{\bf S}_1^z]$ comes from the periodic boundary condition, and evaluates to $-1$ on a site that has half-integer spin. Additionally, since we assume the ground state $|\psi_0\rangle$ is unique, it is an eigenstate of $\sum_{n=1}^L S^z$ with eigenvalue $0$. By combining these, 

\begin{equation} 
	\hat{T}\hat{U}^{\mathrm{twist}}\hat{T}^{-1}|\psi_0\rangle=-\hat{U}^{\mathrm{twist}}|\psi_0\rangle.
 \label{anti-relation}
\end{equation}
By the same uniqueness, $|\psi_0\rangle$ is a momentum eigenstate. This indicates that $|\psi_1\rangle$ has a finite momentum shift of $\pi$ relative to $|\psi_0\rangle$ and is therefore orthogonal to $|\psi_0\rangle$.

Next we calculate the energy difference between $|\psi_1\rangle$ and the ground state $|\psi_0\rangle$:
\begin{equation}
	\begin{split}
		\Delta =& \langle \psi_0|\left[(\hat{U}^{\text{twist}})^\dagger\hat{H}\hat{U}^{\text{twist}}-\hat{H}\right]|\psi_0\rangle\\
		\le& \langle \psi_0|\left[(\hat{U}^{\text{twist}})^\dagger\hat{H}\hat{U}^{\text{twist}}+\hat{U}^{\text{twist}}\hat{H}(\hat{U}^{\text{twist}})^\dagger-2\hat{H}\right]|\psi_0\rangle\\
	\end{split}
\end{equation}
where  we have used the ground state of $\hat{H}$ satisfies $\langle \psi_0|\hat{U}^{\text{twist}}\hat{H}(\hat{U}^{\text{twist}})^\dagger-\hat{H}|\psi_0\rangle \ge 0$. In the Hamiltonian, the $\hat{\bf S}^z \hat{\bf S}^z$ terms are invariant under the spin twist operation, and only the $\hat{\bf S}^+ \hat{\bf S}^-$ and $\hat{\bf S}^- \hat{\bf S}^+$ terms change under the spin twist operation. Using the relation $e^{i \theta \hat{\bf S}_z}\hat{\bf S}^{+}e^{-i\theta \hat{\bf S}_z}=e^{i\theta}\hat{\bf S}^+$, $\hat{\bf S}^+ \hat{\bf S}^-$ transforms under the twist as
\begin{equation}
    \begin{split}
        \hat{U}^{\text{twist}}\hat{\bf S}^{+}_j\hat{\bf S}^{-}_{j+1}(\hat{U}^{\text{twist}})^\dagger &=e^{i\frac{2\pi (r_j-r_{j+1})}{L}}\hat{\bf S}^{+}_j\hat{\bf S}^{-}_{j+1}\\
        &=e^{-i\frac{2\pi }{L}}\hat{\bf S}^{+}_j\hat{\bf S}^{-}_{j+1}.
     \end{split}
\end{equation}

Adding the contribution from the Hermitian conjugates gives an energy difference
\begin{widetext}
\begin{equation}
	\begin{split}
		&|\langle\psi_0|\hat{U}^{\text{twist}}(\hat{\bf S}^{+}_j\hat{\bf S}^{-}_{j+1}+\hat{\bf S}^{-}_j\hat{\bf S}^{+}_{j+1})(\hat{U}^{\text{twist}})^\dagger-\hat{\bf S}^{+}_j\hat{\bf S}^{-}_{j+1}-\hat{\bf S}^{-}_j\hat{\bf S}^{+}_{j+1}|\psi_0\rangle|=2\left[1-\cos(\frac{2\pi }{L})\right]|\langle \psi_0|\hat{\bf S}^{+}_j\hat{\bf S}^{-}_{j+1}|\psi_0\rangle|.
	\end{split}
 \label{E_difference}
\end{equation}
\end{widetext}
Since $\hat{\bf S}^+ {\bf S}^-$ is a bounded operator, this energy difference in Eq.~\eqref{E_difference} is of order $\mathcal O(\frac{1}{L})^2$ in the limit $L\to \infty$. Summing all the $\mathcal O(L)$ terms like this in the Hamiltonian, we obtain that the total energy difference between $|\psi_1\rangle$ and $|\psi_0\rangle$ is proportional to $\frac{1}{L}$.

This shows that if a system satisfying the conditions of the LSM theorem has a unique ground state, we can always find an excited state orthogonal to it without a finite gap. That is, in the thermodynamic limit $L\to \infty$, the energy difference between these states is proportional to $\frac 1L$. This proves the celebrated LSM theorem.

\section{The spectrum gap calculation for long-range Hamiltonion}

In the original proof of the LSM theorem, the locality of the Hamiltonian is important because it limits the number of terms in the Hamiltonian to be linear in the system size $L$. However, for a generic long-range interacting system with many-body interactions, this assumption fails because the Hamiltonian will generally contain exponentially many terms in $L$. A way to circumvent this issue is to only consider $n$-body interactions where $n$ does not scale with $L$. For concreteness, we focus on two-body interactions, whose decay we assume to be bounded by a power law.

Preserving the $U(1)$ rotation symmetry around the $z$-axis, our example in the last section can be generalized to
\begin{equation} 
	\hat{H}=\sum _{m\neq n} \left[C_{mn}\hat{\bf S}_{m}^{+}\hat{\bf S}_n^{-}+D_{mn}\hat{\bf S}_{m}^{z}\hat{\bf S}_n^{z}\right].
\end{equation}
Since $\hat{H}$ is hermitian, $C_{mn}=C_{nm}^*$. Here, we only need $C_{mn}$ to decay no slower than $|m-n|^{-\alpha}$ at the large distance, which contains exponential decay as a trivial case. For the ease of notation below we write $|C_{mn}|<\frac C{\text{min}(|m-n|,L-|m-n|)^\alpha}$ (for periodic boundary condition) with $C$ being an $\mathcal O(1)$ constant. 

Following the original proof, we will show that on a periodic lattice of length $L$, the lowest excitation energy relative to the ground state $|\psi_0\rangle$ is bounded by order $\mathcal O(L^{2-\alpha})$. To do this, we calculate the energy difference between the twisted state $\hat U_\text{twist}$ and the ground state, which upper bounds the desired gap,

%We define the critical distance $r_c$ from $e^{-\mu r_c}=L^{-\alpha}$, Thus, $r_c=\frac{\alpha}{\mu}\log(L)$. This implies that at distances larger than $r_c$, the long-range interaction decays faster than a power-law with decay exponent $\alpha$. Therefore, we obtain
\begin{widetext}
\begin{equation} 
	\begin{split}
        \langle \psi_0| \left[(\hat{U}^{\text{twist}})^\dagger \hat{H}\hat{U}^{\mathrm{twist}} -\hat{H}\right]|\psi_0 \rangle 
        \leq &\langle \psi_0| \left[(\hat{U}^{\text{twist}})^\dagger \hat{H}\hat{U}^{\mathrm{twist}}  +\hat{U}^{\mathrm{twist}}\hat{H}(\hat{U}^{\text{twist}})^\dagger-2\hat{H}\right]|\psi_0 \rangle\\
        =&L \sum_{r=1}^{L-1}C_{1,1+r}\langle \psi_0| \left[(\hat{U}^{\text{twist}})^\dagger\hat{\bf S}_{1}^{+}\hat{\bf S}_{1+r}^{-}\hat{U}^{\mathrm{twist}} +\hat{U}^{\mathrm{twist}}\hat{\bf S}_{1}^{+}\hat{\bf S}_{1+r}(\hat{U}^{\text{twist}})^\dagger-2\hat{\bf S}_{1}^{+}\hat{\bf S}_{1+r}^{-}\right]|\psi_0 \rangle\\
		\le &2L \sum_{r=1}^{L-1}|C_{1,1+r}|\left[1-\cos(\frac{2\pi r}{L})\right]|\langle \psi_0| \hat{\bf S}_{1}^{+}\hat{\bf S}_{1+r}^{-}|\psi_0 \rangle|.
	\end{split}
\end{equation}
\end{widetext}

We will prove that this energy gap is bounded by $O(L^{2-\alpha})$, and thus it vanishes in the limit $L \to \infty$ for $\alpha>2$. The proof idea is straightforward: since $\hat{S}_{1}^{+}\hat{S}_{1+r}^{-}$ is bounded, we only need that for any $r$,
\begin{equation}
|C_{1,1+r}|\left[1-\cos(\frac{2\pi r}L)\right] \leq \mathcal O(L^{-\alpha}).
\end{equation}
%any $r$, $L|C_{1,1+r}|\left[1-\cos(\frac{2\pi r}L)\right]$ cannot exceed order $\mathcal O(L^{1-\alpha})$. 
To show this, we fix a small $k>0$ and separate the problem into a short-distance part where $d\equiv\text{min}(r, L-r)\le kL$ and a long-distance part where $d>kL$. For the short-distance part,
\begin{equation}
\begin{split}
        &|C_{1,1+r}|\left[1-\cos(\frac{2\pi r}L)\right]
        <\frac C {d^\alpha} \left[1-\cos(\frac{2\pi r}L)\right]\\
        =&2\pi^2C\frac{d^{2-\alpha}}{L^2}\left[1+\mathcal O(\frac{d^2}{L^2})\right]\le2\pi^2C\frac{d^{2-\alpha}}{L^2}\left[1+\mathcal O(k^2)\right]
\end{split}
\end{equation}
which is at most of order $\mathcal O(L^{-\alpha})$ for large $L$ and small $d \leq kL$. On the other hand, for the long-distance part where $d>kL$,
\begin{equation}
    |C_{1,1+r}|\left[1-\cos(\frac{2\pi r}L)\right] \leq \frac{2C}{L^{\alpha} k^\alpha},
\end{equation}
which is again of order $\mathcal O(L^{-\alpha})$. Now integrating against $r$ gives the desired bound $\mathcal O(L^{2-\alpha})$.

%On the other hand, the sum of terms with $r > r_c$ contributes 
% 	\begin{equation} 
% 	\begin{split}
% 		 &2L \sum_{r_c\leq r \leq L-1}r^{-\alpha} \left[1-\cos(\frac{2\pi r}{L})\right]|\langle \psi_0| \hat{S}_{1}^{+}\hat{S}_{1+r}^{-}|\psi_0 \rangle|\\
% 		&<4L \sum_{r_c\leq r \leq L-1}r^{-\alpha} |\langle \psi_0| \hat{S}_{1}^{+}\hat{S}_{1+r}^{-}|\psi_0 \rangle|\\
% 		&<4L (L-r_c-1)L^{-\alpha}|\langle \psi_0| \hat{S}_{1}^{+}\hat{S}_{1+r}^{-}|\psi_0 \rangle|\sim L^{2-\alpha}
% 	\end{split}
% \end{equation}
 
 % In the last line, we have used that $\hat{S}_{1}^{+}\hat{S}_{1+r}^{-}$ is a bounded operator. Combing these two parts of contribution together, we have demonstrated that the low-energy excitation gap of the Hamiltonian is bounded by $\max\lbrace\frac{[\log(L)]^3}{L}, L^{2-\alpha}\rbrace$, which vanishes as $L\to \infty$ provided that $\alpha>2$.

From the above analysis, we conclude that in a long-range interacting system with power-law decaying interaction $|m-n|^{-\alpha}$ and $\alpha>2$ at large distances, the spectral gap vanishes at infinite system size. 

Therefore, by following a similar approach to the original proof of the LSM theorem (where the proofs of steps 1 and 2 are the same), we can show that a system with long-range interactions, featuring a power-law decaying interaction with decay exponent $\alpha>2$, also satisfies the LSM theorem.

\section{Numerical results}
Below, we provide some numerical results for two long-range model that satisfy the LSM constraints. The physical Hamiltonian of our first example is a long-range spin-1/2 Heisenberg antiferromagnetic chain model. Its Hamiltonian is written as follows:
\begin{equation}
\hat{H}=\sum_{i\neq j}^L  J_{ij}\hat{\bf{S}}_{i}\cdot \hat{\bf S}_{j}.
\label{heisenberg}
\end{equation}
Here, $J_{ij}=d_{ij}^{-\alpha}$ with $d_{ij}\equiv\text{min}(|i-j|,L-|i-j|)$. The usual spin-1/2 Heisenberg antiferromagnetic chain model satisfies the LSM constraints and has gapless excitation. We expect its long-range interaction version's ground state to be gapless when the decay exponent $\alpha > 2$. 

Below we provide numerical results on the spectral properties of this model with both periodic boundary condition (PBC), where the LSM theorem can be applied, and open boundary condition (OBC) for reference. In the PBC case, numerics show that the gap vanishes for all $\alpha$ we considered, regardless of if $\alpha>2$. In the OBC case, the gap vanishes for $\alpha>2$ and becomes finite for some $\alpha<2$.

\begin{figure}[t]
    \centering
    \includegraphics[width=0.5\textwidth]{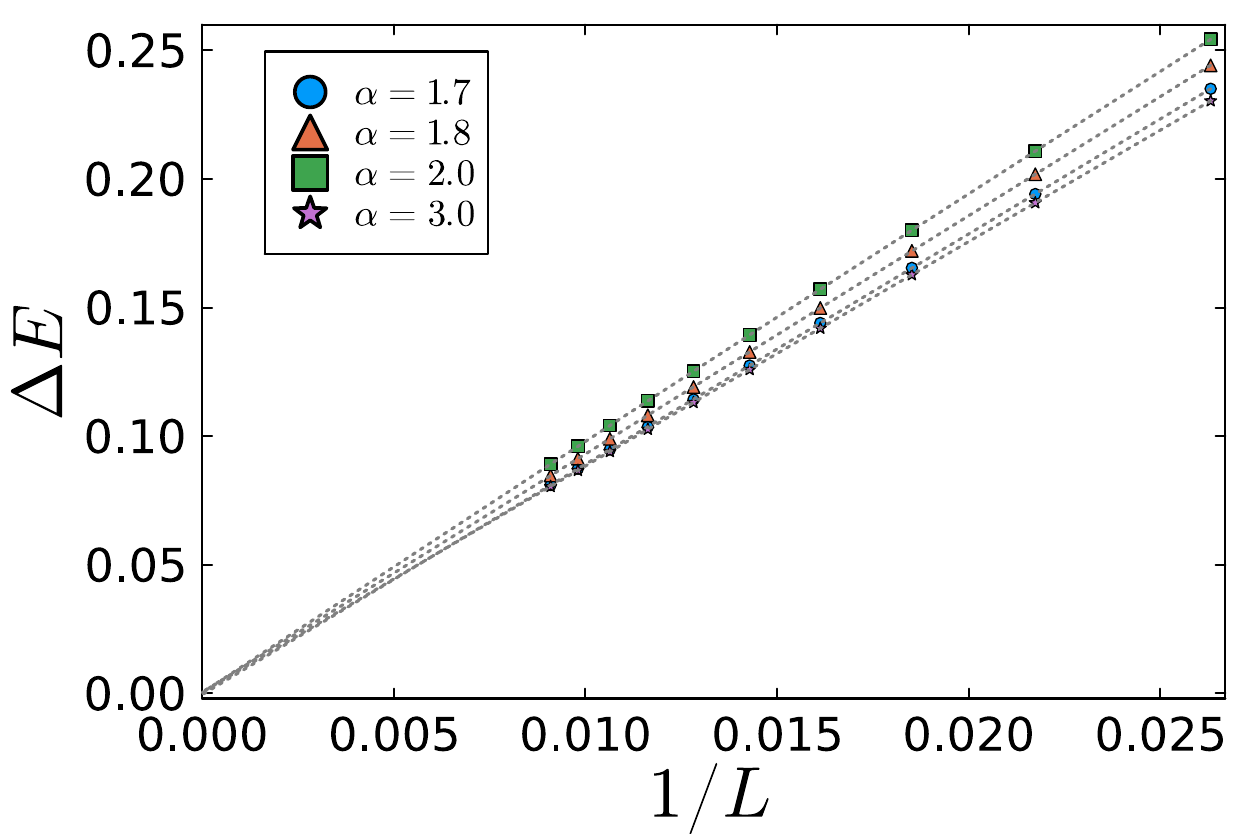}
    \caption{The gap above the ground state of the long-range Heisenberg model is plotted as a function of the total system length $L$ for different values of $\alpha$. The dotted lines are fitting curves with the form $c_0 + c_1/L + c_2/L^2$. Here, we chose the PBC.
    \label{fig:PBC}
    }
\end{figure}

\begin{figure}[t]
    \centering
    \includegraphics[width=0.5\textwidth]{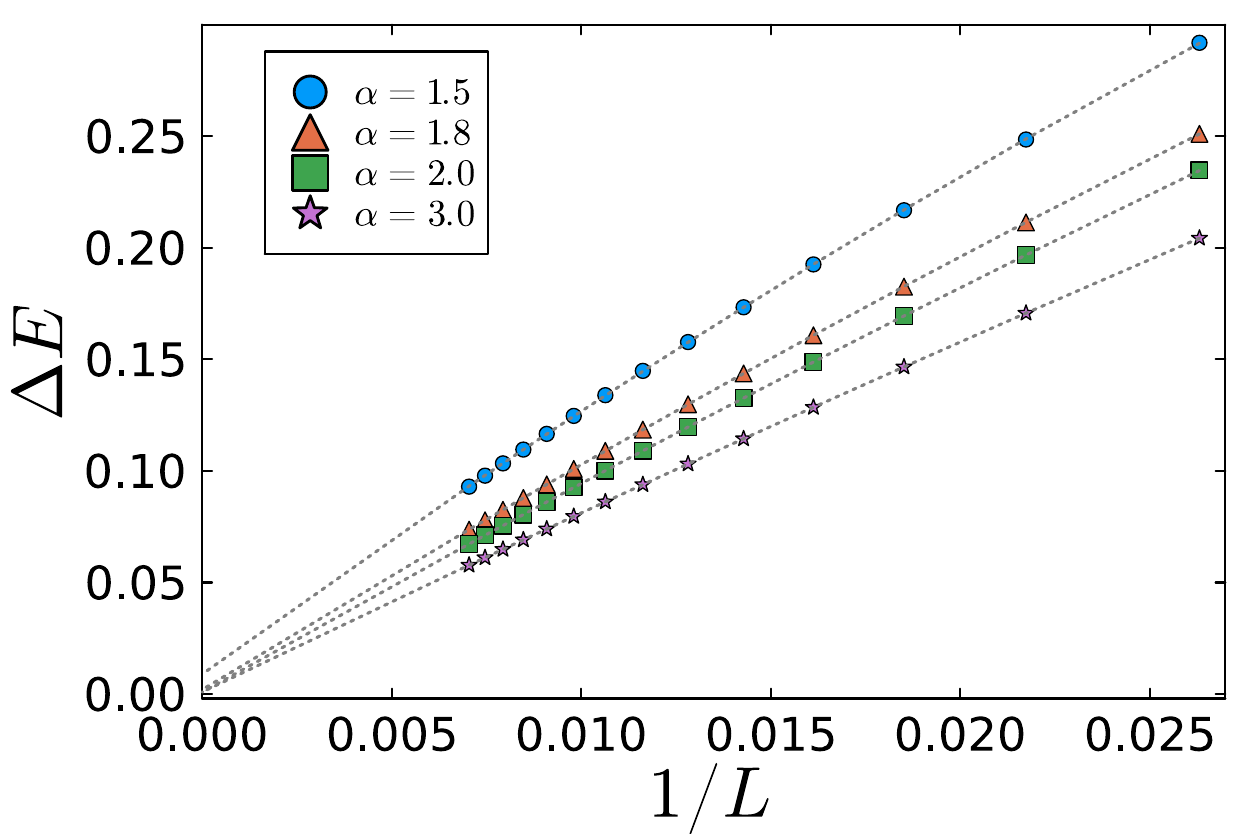}
    \caption{The gap above the ground state of the long-range Heisenberg model is plotted as a function of the total system length $L$ for different values of $\alpha$. The dotted lines are fitting curves with the form $c_0 + c_1/L + c_2/L^2$. Here, we chose the OBC.
    \label{fig:OBC}
    }
\end{figure}
In more detail, we calculate the excitation gap for this model in Eq.~\eqref{heisenberg} and fit the gap against $c_0+c_1/L+c_2/L^2$. We demonstrate the numerical results Fig.~\ref{fig:PBC} for the periodic boundary condition \footnote[2]{The fitting results for the dot line are $(c_0,c_1,c_2)=(-0.0005,8.9648,-0.5098)$ for $\alpha = 1.7$, $(0.0004, 9.2631,0.0032)$ for $\alpha = 1.8$, $(0.0002,9.8339,-6.9200)$ for $\alpha = 2.0$ and $(0.00049,8.8260,-3.4897)$ for $\alpha = 3.0$ in Fig.~\ref{fig:PBC}.}. The corresponding results with open boundary condition are given in Fig.~\ref{fig:OBC} \footnote[1]{The fitting results for the dotted line are $(c_0,c_1,c_2)=(0.0089,12.4089,-63.27)$ for $\alpha = 1.5$, $(0.0020,10.4559,-37.763)$ for $\alpha = 1.8$, $(0.00085,9.6312,-28.485)$ for $\alpha = 2.0$ and $(0.00091,8.2007,-17.9664)$ for $\alpha = 3.0$ in Fig.~\ref{fig:OBC}.}. The fitting curves are depicted as the gray dashed lines in Fig.~\ref{fig:PBC} and Fig.~\ref{fig:OBC}. As shown in Fig.~\ref{fig:PBC} and Fig.~\ref{fig:OBC}, the finite size scaling supports the statement that the spectral gap $\Delta E$ vanishes in the thermodynamic limit $L \rightarrow \infty$ for $\alpha = 2$ and $\alpha = 3$, whereas a finite gap exists for $\alpha = 1.5$ for the open boundary condition depicted as the blue circle line in the Fig.~\ref{fig:OBC}.

% \footnote[1]{The fitting results are 
%     $(c_0,c_1,c_2)=(0.1696,8.52,-24.5)$ ,$(0.0127,11.9,-48.7)$,$(0.0017,9.5,-25.2)$ and $(0.0014,8.13,-15.9)$ for $\alpha = 1.0,1.5,2.0,3.0$ respectively for Fig.~\ref{fig:PBC}.}

% \begin{figure}[t]
%     \centering
%     \includegraphics[width=0.7\textwidth]{Heisenberg_PBC_1.0.pdf}
%     \caption{ The gap above the gap state of the long-range Heisenberg model is plotted as a function of the total system length $L$ for different $\alpha$. The dotted lines are fitting curves with $c_0+c_1/L+c_2/L^2$.Here, we choose the periodic boundary condition.
%     \label{fig:PBC}
%     }
% \end{figure}

The second model we consider is a spin-1/2 Majumdar--Ghosh (MG) model~\cite{Majumdar1969} with long-range interaction. The original MG model reads
\begin{equation}\label{eq:mg}
\hat{H}=\sum_{i=1}^L ~~J_1\left(\hat{\bf{S}}_{i}\cdot \hat{\bf S}_{i+1}+\frac{1}{2}\hat{\bf{S}}_{i}\cdot \hat{\bf {S}}_{i+2}\right).
\end{equation}
 This model has two degenerate ground states in periodic boundary conditions, where neighboring pairs of spins form singlet configurations. 
 
 We modify the original MG model to allow for long-range interactions
\begin{equation}\label{eq:mg_long}
\hat{H}=\sum_{i\neq j}^{L} J_{1,ij}\hat{\bf{S}}_{i}\cdot \hat{\bf S}_{j}+\sum_{i\neq j,j\pm 1}^{L} \frac{1}{2}J_{2,ij}\hat{\bf{S}}_{i}\cdot \hat{\bf {S}}_{j}.
\end{equation}
Here, $J_{1,ij}=|d_{ij}|^{-\alpha}$ and $J_{2,ij}=|d_{ij}-1|^{-\alpha}$ with $d_{ij}\equiv\text{min}(|i-j|,L-|i-j|)$. When $\alpha \to \infty$, this model becomes the original MG model. We expect its ground state to exhibit spontaneous symmetry breaking with periodic boundary condition, provided that the decay exponent $\alpha > 2$. 
\begin{figure}[t]
    \centering
    \includegraphics[width=0.48\textwidth]{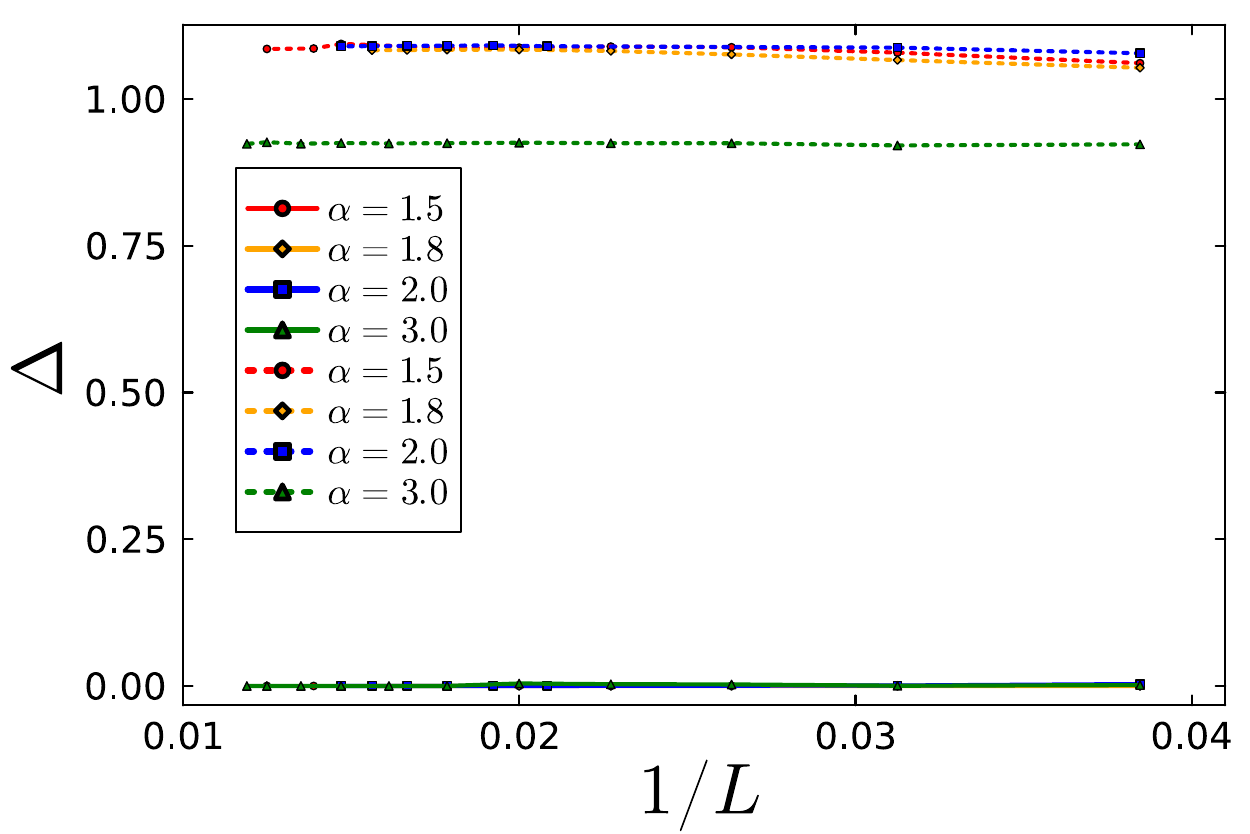}
    \caption{The spectra properties for the long-range MG model defined in Eq.~\eqref{eq:mg_long}. The gap of the first excited state above the ground state $\Delta_1 = E_1-E_0$ (depicted as solid lines) and the second excited state above the ground state $\Delta_2 = E_2-E_0$ (depicted as dot lines) as a function of system size $L$ for different values of $\alpha$. Here, we chose the PBC.% The gray dotted line represents the $0$ value for comparison. (b) The ratio between the gap of the first excited state above the ground state $\Delta_1 = E_1-E_0$ and the second excited state above the ground state $\Delta_2 = E_2-E_0$ as a function of $L$ for different values of $\alpha$. The gray dotted line represents the $0$ value for comparison. The lowest three eigenvalues $E_0$ (blue solid line), $E_1$ (red dash line) and $E_2$ (green dot line) as a function of $L$ for $\alpha=3.0$ in (c) and for $\alpha=1.0$ in (d). 
    \label{fig:MG_all}
    }
\end{figure}

In Fig.~\ref{fig:MG_all}, we plot the gap of the first excited state above the ground state $\Delta_1 = E_1-E_0$ (depicted as the solid line) and the second excited state above the ground state $\Delta_2 = E_2-E_0$ (depicted as the dot line) as a function of system size $L$ for different values of $\alpha=1.5,1.8,2,3$ for the long-range MG model with PBC. Here, $E_0, E_1$, and $E_2$ denote the energy of the ground state, the first excited state, and the second excited state respectively. We find that these numerical results suggest the double degeneracy of ground state always exists for $\alpha\ge2$ (see the blue line for $\alpha=2$ and the green line for $\alpha=3$). Our numerics show that as in the Heisengerb example, this degeneracy persists for $\alpha<2$. %Our data is more inconclusive for the $\alpha=1.5,$ case. 

\begin{figure}[t]
    \centering
    \includegraphics[width=0.48 \textwidth]{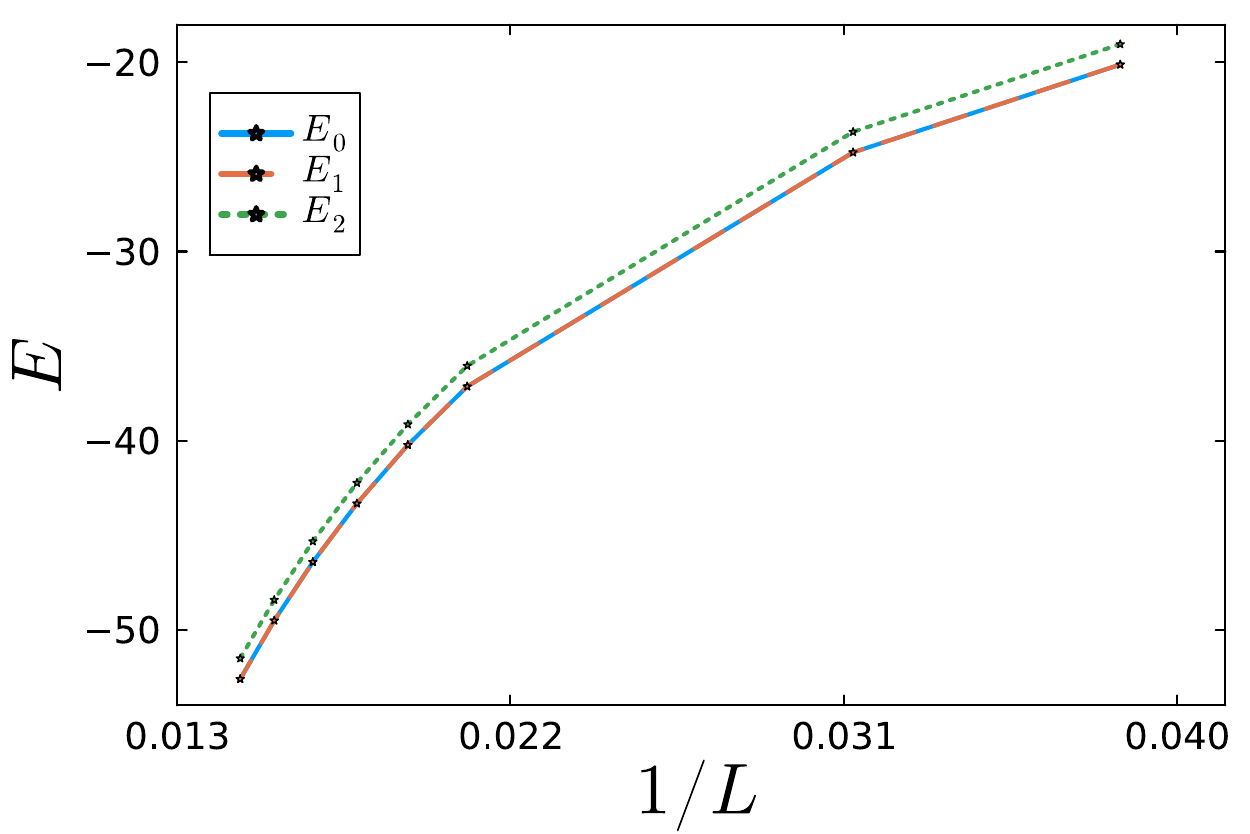}
    \caption{The lowest three eigenvalues $E_0$, $E_1$ , and $E_2$ of the long-range MG model defined in Eq.~\eqref{eq:mg_long} as a function of $L$ for $\alpha=2.0$. We use periodic boundary conditions in our numerical analysis. Here, we chose the PBC.
    \label{fig:MG_2.0}
}
\end{figure}

We also plot the lowest three eigenvalues $E_0$ (blue solid line), $E_1$ (red dashed line), and $E_2$ (green dotted line)) of the long-range MG model as a function of the total system length $L$ for $\alpha=2.0$ in Fig.~\ref{fig:MG_2.0}. The double-degenerate ground state is visible for $\alpha=2.0$ within the system size we simulated.

\section{Summary and Outlook}
In the present work, we consider the fate of the LSM theorem in systems with long-range interactions. We find that for systems satisfying LSM constraints with either exponential decay or power-law decay with a decay exponent $\alpha > 2$, the LSM implications on the ground state properties still hold. This means that their ground state is either gapless or gapped with degeneracy. We also provide numerical evidence for the long-range versions of the Heisenberg model and the Majumdar–Ghosh model to verify our findings. In these two models, changing the original short-range interaction to a long-range interaction with a decay exponent $\alpha > 2$ does not alter their gapless excitation or degenerate ground state. However, for $\alpha < 2$, the ground state could become trivially gapped.

This convenient transition point $\alpha=2$ makes it possible to observe the LSM constraint on ground state properties in real experiments. For example, on the trapped ion platform, the approximate power-law exponent can be tuned within the range $0 < \alpha < 3$ by adjusting the detuning of the laser fields from the vibrational sidebands and by varying the intensity of the lasers and the trap frequencies~\cite{Britton_2012}. Consequently, it is possible to witness the breakdown of the LSM constraints on the ground state as one tunes the decay exponent from $\alpha > 2$ to $\alpha < 2$. Additionally, it will be interesting to explore the relationship between the breakdown of the LSM theorem in the long-range interacting system and the quantum phase transition, the critical properties, and the dynamical behavior of physical observables in the quantum systems as we tune the decay exponent.

\section*{Acknowledgements}
We thank Chengshu Li, Yingfei Gu, and Hui Zhai for the helpful discussions in our previous work about the LSM theorem.  The DMRG calculations are performed using the ITensor library (v0.3) \cite{ITensor}.

\section*{Note added}

We obtained this long-range version of the LSM theorem during our previous work about the LSM theorem. Recently, we have noticed increasing interest in long-range interaction systems, which led us to finally summarize our results in this paper. There are two other independently noteworthy papers that have been studying similar LSM constraints in long-range models \cite{liu2024liebschultzmattis,ma2024liebschultzmattis}.

\bibliography{ref.bib}% Produces the bibliography via BibTeX.

\end{document}